\documentstyle[preprint,aps]{revtex}
\preprint{USM-TH-82}
\begin{document}
\title{From Screening to Confinement in a Gauge-Invariant Formalism}
\author{ Patricio Gaete \thanks{E-mail: pgaete@fis.utfsm.cl},
and Iv\'an Schmidt \thanks{E-mail: ischmidt@fis.utfsm.cl}}
\address{Departamento de F\'{\i}sica, Universidad T\'ecnica F.
Santa Maria, Valpara\'{\i}so, Chile}
\maketitle

\begin{abstract}
Features of screening and confinement are reviewed in
two-dimensional quantum electrodynamics (QED2). Our discussion is
carried out using the gauge-invariant but path-dependent variables
formalism. This alternative and useful approach exploits the rich
structure of the electromagnetic cloud or dressing around static
fermions in a straightforward and simple way.
\end{abstract}
\smallskip

PACS number(s): 12.20.Ds, 11.15.Tk

\section{INTRODUCTION}

It is well known that one of the long standing problems in physics
is understanding confinement of quarks and gluons from first
principles. The reason is that infrared divergences and gauge
dependence make bound-state equations very difficult to
approximate. In this paper we want to address the issue of gauge
dependence within the confinement problem. Our purpose is to
present a formalism in which everything is expressed in terms of
physical (gauge invariant) quantities from the start. As a bonus,
the usual qualitative picture of confinement, in terms of an
electric flux tube linking quarks \cite{Muta,Cheng}, emerges
naturally in this formalism.

As a first step we will consider two dimensional gauge theories,
which can be regarded as a theoretical laboratory for
studying four-dimensional gauge theories. Of particular interest
are non-perturbative issues such as confinement and spectrum
of models, which can be settled in these theories. Of these, the
Schwinger model\cite{Schmod} has probably enjoyed the greatest
popularity due to several features that it possesses. For
instance, the spectrum contains a massive mode, the charge is
screened and confinement is satisfactorily addressed. We also
draw attention to the fact that the transition from screening to
confinement of probe charges is possible only for nonvanishing
spinor mass\cite{Gross,Elcio}.

On the other hand, in recent times a description in terms of
gauge-invariant but path-dependent field variables
in Abelian gauge theories, and the intimately related question of
gauge fixing, has been developed\cite{Gaete1,Gaete2}. In
particular, it was shown how the gauge fixing procedure
corresponds, in this formalism, to a path choice. Therefore this
represents a path-dependent but physical QED where a consistent
quantization directly in the path space is carried out.
Incidentally, it is of interest to notice that the physical
electron is not the Lagrangian fermion, which is neither
gauge-invariant nor associated with an electric field. Instead,
the physical electron is the Lagrangian fermion together with a
non-local cloud (or dressing) of gauge fields\cite{Dirac}. This
remark opens up the way to a stimulating discussion of how the
electromagnetic cloud is distributed around fermions.

Within this framework the aim of the present paper is to
reexamine some aspects about screening and confinement in
two-dimensional quantum electrodynamics from the viewpoint of the
gauge-invariant formalism. This offers
a natural setting for such studies, because it involves the use of
strings to carry electric flux. Moreover, we obtain computational
rules that have clear as well as simple interpretation, in
contrast to the standard Wilson loop procedure where subtleties
related to the correct calculation must be
considered\cite{Haagen,Fidel}. In Sec. II we present
gauge-invariant expressions which will form the basis of our
subsequent considerations. Sec. III constitutes the central part
of our work. Specifically, we calculate
the interaction energy between external probe sources, paying due
attention to the structure of the fields that surround the
charges. Here we will focus our attention on the transition from
screening to confinement.

\section{GAUGE-INVARIANT \ VARIABLES}

Let us start our analysis with a brief presentation of the
gauge-invariant variables formalism \cite{Gaete1,Gaete2}. For this purpose,
we introduce the vector gauge-invariant field
\begin{equation}
{\cal A}_{\mu }(y)=A_{\mu }(y)+\partial _{\mu }\left( -\int_{C_{\xi
y}}dz^{\mu }A_{\mu }(z)\right) ,  \label{bia}
\end{equation}
where the path integral is to be evaluated along some contour $C_{\xi y\text{ }}$%
connecting $\xi $ and $y$. Here $A_{\mu }$ is the usual
electromagnetic potential and, in principle, it is taken in an
arbitrary gauge. The point we wish to emphasize, however, is that
${\cal A}_{\mu }(y)$ is invariant with respect to gauge
transformations \begin{equation}
A_{\mu} (y)\rightarrow A_{\mu} (y)+\partial _{\mu}\Lambda (y).
\label{vecta}
\end{equation}
Thus ${\cal A}_{\mu }$, while no longer gauge-dependent, now
becomes path-dependent. We now
choose the contour as the spacelike straight line $ z^1  = \xi ^1
+ \alpha (y - \xi )^1 $, where $\alpha $ $ (0$ $\leq \alpha $
$\leq 1)$ is the parameter describing the contour and $ \xi^1 $ is an
arbitrary (fixed) reference point. Without loss of generality, we
can choose $ \xi^1 $ = 0. This path choice may be made more
explicit by splitting Eq.(\ref{bia}) in the form
\begin{equation}
{\cal A}_1 (x) = A_1 (x) - \partial _1 \int\limits_0^1 {d\alpha }
x^1 A_1 (\alpha x^1 ) ,
\end{equation}  \label{bea1}
\begin{equation}
{\cal A}_0 (x) = A_0 (x) - \partial _0 \int\limits_0^1 {d\alpha
x^1 } A_1 (\alpha x^1 ) , \label{bea2}
\end{equation}
a short calculation yields
\begin{equation}
{\cal A}_1 (x^0 ,x^1 ) = 0 , \label{bea3}
\end{equation}
\begin{equation}
{\cal A}_0 (x^0 ,x^1 ) = \int\limits_0^1 {d\alpha x^1 } E_1
(\alpha x^1 ) , \label{bea4}
\end{equation}
where $ E_1 $ is the one-dimensional electric field. The above
expressions coincide with the Poincar\'{e} gauge conditions
\cite{Gaete1}.

We now turn our attention to the fermion field. In this formalism
the charged matter field together with the electromagnetic cloud
(dressing) which surrounds it, or what is the same the physical
electron, is given by \cite{Gaete1,Dirac},
\begin{equation}
\Psi (y) = \exp \left( { - ie\int_{C_{\xi y} } {dz^\mu  A_\mu (z)}
} \right)\psi (y) . \label{expo1}
\end{equation}
Following our path choice, expression ( \ref{expo1}) may be
rewritten as
\begin{equation}
\Psi (y) = \exp \left( { - ie\int_\xi ^y {dz^1 A_1 (z)} }
\right)\psi (y) . \label{expo2}
\end{equation}
It is worthwhile remarking at this point that the breaking of the
gauge invariance of the fields in the standard formalism is
transformed into breaking of the translational invariance in the
path-dependent formalism. The way of solving this problem is
letting the reference point $ \xi^1 $ go to infinity.

Before we proceed further, we wish to show that this
approach yields interesting results by calculating a
gauge-invariant photon propagator. From the expression for the
physical fields (\ref{bia}) it can be shown that the
gauge-invariant propagator is given by \begin{eqnarray} {\cal
D}_{\mu \nu } (x,y) = D_{\mu \nu } (x,y) + \frac{\partial
}{{\partial x^\mu  }}\int_x^\xi {dz^\alpha  } D_{\alpha \nu }
(z,y) + \frac{\partial }{{\partial y^\nu }}\int_y^\xi  {dw^\beta
D_{\mu \beta } (x,w)  + } \nonumber \\  + \frac{\partial
}{{\partial x^\mu \partial y^\nu  }} \int_x^\xi  {dz^\alpha  }
\int_y^\xi  {dw^\beta  D_{\alpha \beta } (z,w)} . \label{prop1}
\end{eqnarray}
$D_{\mu \nu }(x,y)$ on the right-hand side (RHS) of Eq.(\ref{prop1}) is the
photon propagator taken in an arbitrary gauge. Implementation of
the Poincar\'{e}  gauge amounts to using the contour
$z^1=\xi^1 + \alpha (y-\xi)^1$  and  $z^1=\xi^1 +\rho
(x-\xi)^1 $. We can choose, for example, $D_{\mu \nu } (x,y)$ in
the temporal gauge, that is,
\begin{equation}
D_{\mu \nu } (x,y) = g_{\mu 1} g_{\nu 1}   \delta (x^1  - y^1
)\left( {\frac{1}{2}|x^0  - y^0 | + B(x^0  - y^0 ) - A} \right)
,\label{prop2}
\end{equation}
where, as it is well known, the residual gauge invariance
manifests itself in the presence of the constants $A$ and $B$.
This is a peculiarity of the temporal gauge, which does not fix
the gauge uniquely. Then, from Eq.(\ref{prop1}), we find that
\begin{equation}
{\cal D}_{\mu \nu } (x,y) = g_{\mu 0} g_{\nu 0} \frac{1}{2}\delta
(x^0 - y^0 )\left( {|x^1  - y^1 | - |x^1  - \xi ^1 | - |y^1  - \xi
^1 |} \right) . \label{prop3}
\end{equation}
Thus one avoids the unphysical features associated with the
gauge-dependent formulation. Nevertheless, as it should be
expected, the above propagator breaks the translational
invariance. The solution to this problem is letting $\xi^1$
go to infinity. However, we do not intend to address these
problems here. A fuller account on gauge invariant Green's
functions will be provided elsewhere \cite{Gaete3}.

This concludes our brief introduction to gauge invariant variables
and gauge conditions.

\section{INTERACTION ENERGY IN QED2}
As already mentioned, our immediate objective is to calculate the
interaction energy between external probe sources in the Schwinger
model. To do this, we will exploit the rich structure of the
electromagnetic cloud or dressing around static fermions.

We shall begin by considering the bosonized form of the Schwinger
model \cite{Suss}:
\begin{equation}
{\cal L} =  - \frac{1}{4}F_{\mu \nu }^2  + \frac{1}{2}(\partial
_\mu \varphi )(\partial ^\mu  \varphi ) - \frac{e}{{2\sqrt \pi
}}\varepsilon ^{\mu \nu } F_{\mu \nu } \varphi  +   + m\sum \left(
{\cos \left( {2\pi \varphi  + \theta } \right) - 1} \right) - A_0
J^0 ,   \label{boso}
\end{equation}
where $J^0$ is the external current,  $  \sum  = \frac{e}{{2\pi
^{\frac{3}{2}} }}\exp (\gamma _E ) $ with $\gamma _E $ the
Euler-Mascheroni constant, and $\theta $ refers to the $ \theta $
vacuum.
\subsection {Massless case}
Our purpose is to compute the interaction energy in the $ m = 0 $
case. The first step in this direction is to carry out the
integration over $\varphi $ in (\ref{boso}). This allows us to
write the effective Lagrangian
\begin{equation}
{\cal L} =  - \frac{1}{4}F_{\mu \nu } \left( {1 + \frac{{e^2
}}{\pi }\frac{1}{{\partial ^2 }}} \right)F^{\mu \nu }  - A_0 J^0 .
\label{effec}
\end{equation}
It is worthwhile sketching at this point the canonical
quantization of this theory from the Hamiltonian analysis point of
view. The canonical momenta are $ \Pi ^\mu   =  - \left( {1 +
\frac{{e^2 }}{\pi }\frac{1}{{\partial ^2 }}} \right)F^{0\mu }$
with the only nonvanishing canonical Poisson brackets being
\begin{equation}
\left\{ {A_\mu  (t,x),\Pi ^\nu  (t,y)} \right\} = \delta _\mu ^\nu
\delta (x - y) . \label{pois}
\end{equation}
Since $ \Pi_0 $ vanishes we have the usual primary constraint $
\Pi_0 = 0 $ , and $  \Pi ^1  = \left( {1 + \frac{{e^2 }}{\pi
}\frac{1}{{\partial ^2 }}} \right)F^{10} $ . Therefore the
canonical Hamiltonian is
\begin{equation}
H_C  = \int {dx\left(  { - \frac{1}{2} \Pi _1 \left( {1 +
\frac{{e^2 }}{\pi }\frac{1}{{\partial ^2 }}} \right)^{ - 1}  \Pi
^1 + \Pi ^1
\partial _1 A_0  + A_0 J^0 } \right)} . \label{cano}
\end{equation}
Requiring the primary constraint $\Pi^0 = 0 $ to be preserved in
time yields the following secondary constraint
\begin{equation}
\Omega _1 (x) = \partial _1 \Pi ^1  - J^0 . \label{vinc}
\end{equation}
It is straightforward to check that there are no more constraints
in the theory and that both constraints are first class. The
Hamiltonian that generates translations in time is given by
\begin{equation}
H = H_C  + \int {dx\left( {c_0 (x)\Pi _0 (x) + c_1 (x)\Omega _1
(x)} \right)} , \label{canon}
\end{equation}
where $ c_0(x) $ and $ c_1(x) $ are arbitrary functions.
Furthermore, since $\Pi^0 =0 $ always, and $\stackrel{\bf{\cdot }}
{A_{0}} (x) = \left[ {A_0 (x),H} \right] = c_0 (x)$ , we discard $
A_0(x)$ and $\Pi_0(x)$ . Therefore the Hamiltonian reduces to
\begin{equation}
H = \int {dx\left\{ { - \frac{1}{2}\Pi _1 \left( {1 + \frac{{e^2
}}{\pi }\frac{1}{{\partial ^2 }}} \right) ^{-1} \Pi ^1  +
c^{\prime} (x)\left( {\partial _1 \Pi ^1  - J^0 } \right)}
\right\}} , \label{canon1}
\end{equation}
where $ c^{\prime }(x)=c_{1}(x)-A_{0}(x)$.

According to the usual procedure we introduce a supplementary
condition on the vector potential such that the full set of
constraints becomes second class, so we write
\begin{equation}
\Omega _2 (x) = \int_0^1 {d\alpha x^1 } A_1 (\alpha x) = 0 ,
\label{poin}
\end{equation}
where, as in the previous section, $\alpha$ is the parameter
describing a spacelike straight line of integration. It
immediately follows that the fundamental Dirac brackets read
\begin{equation}
\left\{ {A_1 (x),A^1 (y)} \right\}^ *   = 0 = \left\{ {\Pi _1
(x),\Pi ^1 (x)} \right\}^ * , \label{paren}
\end{equation}
\begin{equation}
\left\{ {A_1 (x),\Pi ^1 (y)} \right\}^ *   = \delta ^{(1)} (x - y)
- \partial _1^x \int_0^1 {d\alpha x^1 \delta ^{(1)} } (\alpha x -
y) . \label{paren2}
\end{equation}
It is important to realize that expression (\ref{expo2}) represents
charged particles together with an associated proper electric field.
To see how this arises let $\mid $ $E\rangle $ be an eigenvector of
the electric field operator $E_{1}(x)$, with eigenvalue
$\varepsilon _{1}(x)$ :
\begin{equation}
E_{1}(x)\mid E\rangle=\varepsilon_{1}(x)\mid E\rangle .
\label{eige}
\end{equation}
Next we will consider the state $\Psi (y)\mid E\rangle $. By means
of Eq. (\ref{eige}) we have that
\begin{equation}
E_1 (x)\text{}\Psi (y)\mid E\rangle  = \Psi (y)\text{}E_{1}
(x)\mid E\rangle +\left[ E_{1} (x),\Psi (y)\right]\mid E\rangle .
\label {eige2}
\end{equation}
From our above hamiltonian analysis, Eq. (\ref{eige2}) may be
rewritten as
\begin{equation}
E_1 (x)\Psi (y)\mid E\rangle  = \left( {\varepsilon _1 (x) +
q\left( {1 + \frac{{e^2 }}{\pi }\frac{1}{{\partial _x^2 }}}
\right)^{ - 1} \int_0^1 {d\alpha y_1 \delta ^{(1)} (\alpha y_1  -
x_1 )} } \right)\Psi (y)\mid E\rangle . \label {eige3}
\end{equation}
Hence we see that the operator $\Psi (y) $ is the dressing
operator of the creation of an electron together with an
associated proper electric field. Notice that the integral in
Eq.(\ref{eige3}) is nonvanishing only on the contour of
integration. As a result, we have a static electric field on a
line.

At this point we should mention that if we consider a modified form
for the electromagnetic cloud in the Poincar\'e gauge Eq.(\ref{expo2}),
which is equivalent to the Coulomb gauge
\cite{Gaete1}, that is,
\begin{equation}
\Psi (y) = \exp \left( { - iq\int_0^y {dz^k A_k^L (z)} }
\right)\psi (y) , \label{coul}
\end{equation}
where $A_{1}^{L}$ refers to the longitudinal part of $A_{1}$, we
would obtain that the field $\Psi $ dresses the charge $\psi $
with the electric field :
\begin{equation}
E_1 (x)\Psi (y)\mid E\rangle  = \left( {\varepsilon _1 (x) +
\frac{q}{2}e^{ - \frac{e}{{\sqrt \pi  }}|x_1  - y_1 |} }
\right)\Psi (y)\mid E\rangle  . \label{coul2}
\end{equation}

In order to calculate the energy between external static charges, we
take a fermion localized at ${y}^{\prime}_1$ and an antifermion at
$ y_1$, both dressed according to Eq.(\ref{expo2}), and compute
the expectation value of the QED2 Hamiltonian in the physical
state $ \mid {\Omega}\rangle $, which we will denote by $\langle
H\rangle_{\Omega}$. From our above Hamiltonian structure, we have
that
\begin{equation}
\left\langle H \right\rangle _\Omega   = \left\langle \Omega
\right|\int {dx_1 } \left( { - \frac{1}{2}\Pi _1 \left( {1 +
\frac{{e^2 }}{\pi }\frac{1}{{\partial ^2 }}} \right)^{-1} \Pi ^1 }
\right)\left| \Omega  \right\rangle  . \label{ener1}
\end{equation}
As mentioned before, the fermions are taken to be static, thus we
can substitute $ \partial^2$ by -$\partial^2_1$ in
Eq.(\ref{ener1}). In that case we write
\begin{equation}
\left\langle H \right\rangle _\Omega  = \left\langle \Omega
\right|\int {dx_1 } \left( { - \frac{1}{2}\Pi _1 \left( {1 -
\frac{{e^2 }}{\pi }\frac{1}{{\partial _1^2 }}} \right)^{-1} \Pi ^1
} \right)\left| \Omega \right\rangle . \label{ener2}
\end{equation}
As has been established by Dirac \cite{Dirac}, the physical states
$ |\Omega\rangle $ correspond to the gauge invariant ones.
In this way, the state corresponding to two opposite charges at
different points can be made gauge invariant by including a dressing
as in Eq.(\ref{expo2}), which keeps the entire state gauge
invariant. In other words,
\begin{equation}
\mid \Omega  \rangle  \equiv \mid \overline{\Psi}
(y)\Psi (y ^{\prime})\rangle  =\overline{\psi }(y)\exp \left( { -
iq\int_y^{y^{\prime}} {dz^1 A_1 (z)} } \right)\psi (y)\left| 0
\right\rangle , \label{ener3}
\end{equation}
where $ | 0\rangle $ is the physical vacuum state.

We are now ready to calculate $\left\langle H\right\rangle _\Omega
$ . Using our formalism, we can show that
\begin{equation}
E_1 (x)\left| \Omega  \right\rangle  = \overline \Psi  (y)\Psi
(y^\prime  )E_1 (x)\left| 0 \right\rangle  + q\left( {1 -
\frac{{e^2 }}{\pi }\frac{1}{{\partial _1^2 }}} \right)^{ - 1}
\int_y^{y^\prime  } {dz_1 \delta (x_1  - z_1 } )\left| \Omega
\right\rangle . \label{inter1}
\end{equation}
Inserting this into Eq. (\ref {ener2}), the energy in the
presence of the static charges will be given by
\begin{equation}
\left\langle H \right\rangle _\Omega   = \left\langle H
\right\rangle _0  + \frac{{q^2 }}{2}\frac{{\sqrt \pi  }}{e}\left(
{1 - e^{ - \frac{e}{{\sqrt \pi  }}|y - y^\prime  |} } \right) ,
\label{inter2}
\end{equation}
where $ \left\langle H \right\rangle _o  = \left\langle 0
\right|H\left| 0 \right\rangle $ . Since the potential is given
by the term of the energy which depends on the separation of the
two fermions, from the expression (\ref{inter2}) we obtain
\begin{equation}
V = \frac{{q^2 }}{2}\frac{{\sqrt \pi  }}{e}\left( {1 - e^{ -
\frac{e}{{\sqrt \pi  }}|y - y^\prime  |} } \right) . \label{pot1}
\end{equation}
Thus we have demonstrated that the potential between fermions can
be directly obtained once the structure of the photonic clouds
around static fermions is known. In this case expression
(\ref{pot1}) is the expected screening contribution to the
potential. Physically this means that the initial string was
broken and all the charges are screened. In other terms, as a
result of the interaction with massless fermions the original
Coulomb potential ( proportional to the distance ) is screened.
This feature is similar to that expected for QCD strings in the
adjoint representation.

Before we conclude this subsection, it is important to notice that
with the path choice stated in Eq.(\ref {coul}) (modified
Poincar\'e gauge), and from the previous canonical formalism, we
can write a scalar potential
\begin{equation}
{\cal A}_o (t,x) = \int_0^1 {d\alpha x^1 } E^L_1 (t,\alpha x) =
\int_0^1 {d\alpha x^1 \left( {1 - \frac{{e^2 }}{\pi
}\frac{1}{{\partial _1^2 }}} \right)} _{\alpha x}^{-1}
\frac{{\partial _1^{\alpha x} \left( { - J^0 (\alpha x)}
\right)}}{{\partial _{\alpha x}^2 }} , \label{comp0}
\end{equation}
where the superscript $L$ refers to the longitudinal part and
$J^0$ is the external source. Accordingly, the potential for a
pair of static pointlike opposite charges located at $y$ and
$y^{\prime}$ , that is, $J^0 (t,x)= q \{
\delta(x-y)-\delta(x-y^{\prime}) \}$, is given by
\begin{equation}
V = q\left( {{\cal A}_0 (y) - {\cal A}_0 (y^\prime  )} \right) =
\frac{{q^2 }}{2}\frac{{\sqrt \pi  }}{e}\left( {1 - e^{ -
\frac{e}{{\sqrt \pi }}|y - y^\prime  |} } \right) . \label{inp}
\end{equation}
It is gratifying to notice here the simplicity and directness of
this derivation, which is manifestly gauge-invariant.
\subsection{Massive case}

We now proceed to consider the massive case. For this purpose we
have to carry out the integration over $ \varphi $ in (\ref{boso}).
But since this expression is non-polynomial in
$\varphi$, we expand the effective Lagrangian in terms of $
F^{\mu \nu}$. Thus it follows that
\begin{equation}
{\cal L} =  - \frac{1}{4}F_{\mu \nu }^2  - \frac{{e^2 }}{{4\pi
}}F_{\mu \nu } \frac{1}{{\partial ^2  + 4\pi m\Sigma }}F^{\mu \nu
}  - A_0 J^0 , \label{bos2}
\end{equation}
where we have taken $\theta=0$. If we now look at the limit
of slowly varying fields, we find that
\begin{equation}
{\cal L} =  - \frac{1}{4}F_{\mu \nu } \left( {1 + \frac{{e^2
}}{{4\pi ^2 m\Sigma }}} \right)F^{\mu \nu }  - A_0 J^0 .
\label{boso3} \end{equation} As in the previous subsection, our
objective will be to calculate the potential energy for this
theory. However, as we now know, this calculation is facilitated
by using the expression:
\begin{equation}
{\cal A}_0 (t,x) = \int_0^1 {d\alpha } x^1 E_1^L (t,\alpha x) .
\label{esc}
\end{equation}
Thus we obtain
\begin{equation}
{\cal A}_0 (t,x) = \left( {1 + \frac{{e^2 }}{{4\pi ^2 m\Sigma }}}
\right)^{ - 1} \int_0^1 {d\alpha } x^1 \frac{{\partial _1^{\alpha
x} ( - J^0 (\alpha x))}}{{\partial _{\alpha x}^2 }} . \label{esc2}
\end{equation}
For $ J^0 (t,\alpha x)= q \delta (\alpha(x-a))$ expression
(\ref{esc2}) then becomes
\begin{equation}
{\cal A}_0 (t,x) = \frac{q}{2}\left( {1 + \frac{{e^2 }}{{4\pi ^2
m\Sigma }}} \right)^{ - 1} |x - a| . \label{esc3}
\end{equation}
By means of expression (\ref{esc3}) we evaluate the interaction
energy for a pair of static pointlike opposite charges at $y$ and
$ y{\prime}$, as
\begin{equation}
V = q\left( {{\cal A}_0 (y) - {\cal A}_0 ( y{\prime})} \right) =
\frac{{q^2 }}{2}\left( {1 + \frac{{e^2 }}{{4\pi ^2 m\Sigma }}}
\right)^{ - 1} |y - y{\prime}| . \label{esc4}
\end{equation}
Considering the limit $m \ll e $, we get
\begin{equation}
V = \frac{{q^2 }}{{e^2 }}2\pi ^2 m\Sigma |y - y{\prime}| .
\label{esc5}
\end{equation}
This can be recognized as the standard result for the interaction
potential \cite{Gross}, which is also just the confinement
contribution to the potential. This derivation tells us that one
can in fact interpolate between screening and confinement as soon
as the dynamical fermions have a nonvanishing mass. It is
therefore of interest to reexamine the transition between these
limits. To see how this arises in this formalism, we start with
Eq.(\ref{bos2}). Using the fact that the fields are taken to be
static, which means substituting $\partial^2$ by $-\partial^2_1$,
we get
\begin{equation}
{\cal L} =  - \frac{1}{4}F_{\mu \nu }^2  + \frac{{e^2 }}{{4\pi
}}F_{\mu \nu } \frac{1}{{\partial _1^2  - 4\pi m\Sigma }}F^{\mu
\nu }  - A_0 J^0 . \label{lag}
\end{equation}

Thus, in the present approach, the scalar potential (\ref{esc})
may be written as
\begin{equation}
{\cal A}_0 (t,x) = \int_0^1 {d\alpha } x^1 \partial _1^{\alpha x}
\left( { - \frac{{J^0 (\alpha x)}}{{\partial _{\alpha x}^2  -
\lambda ^2 }}} \right) - 4\pi m\Sigma \int_0^1 {d\alpha } x^1
\frac{{\partial _1^{\alpha x} ( - J^0 (\alpha x))}}{{(\partial ^2
- \lambda ^2 )_{\alpha x} \partial _{\alpha x}^2 }} , \label{pte}
\end{equation}
where $  \lambda ^2  \equiv \frac{{e^2 }}{\pi } + 4\pi m\Sigma $.
For $ J^0 (t,\alpha x)= q \delta (\alpha (x-a))$  Eq.(\ref{pte})
reduces to
\begin{equation}
{\cal A}_0 (t,x) = \frac{q}{{2\lambda }}\left( {1 + \frac{{4\pi
m\Sigma }}{{\lambda ^2 }}} \right)\left( {1 - e^{ - \lambda a} }
\right) + \frac{q}{2}\left( {1 - \frac{{\frac{{e^2 }}{\pi
}}}{{\lambda ^2 }}} \right)a . \label{pte2}
\end{equation}
Expression (\ref{pte2}) immediately shows that the potential for
two opposite charges located at $ y $ and $ y {\prime}$ is given
by \begin{equation}
V = \frac{{q^2 }}{{2\lambda }}\left( {1 + \frac{{4\pi m\Sigma
}}{{\lambda ^2 }}} \right)\left( {1 - e^{ - \lambda |y -
y{\prime}|} } \right) + \frac{{q^2 }}{2}\left( {1 -
\frac{{\frac{{e^2 }}{\pi }}}{{\lambda ^2 }}} \right)|y -
y{\prime}| . \label{ptb}
\end{equation}
It is straightforward to check that in the limit $ m=0 $,
expression (\ref{ptb}) reduces to (\ref{inp}).

Until now we have taken the vacuum angle $\theta$ as zero. We now
want to consider the nonvanishing $\theta$ contribution to the
potential that would follow from this formalism. With this in
mind, we start by writing
\begin{equation}
{\cal L} = -\frac{1}{4}F_{\mu \nu } F^{\mu \nu }  - \frac{{e^2
}}{{4\pi }}F_{\mu \nu } \frac{1}{{\partial ^2  + 4\pi m\Sigma
}}F^{\mu \nu }  + em\Sigma  \frac{1}{{\partial ^2  + 4\pi m\Sigma
}}\theta \varepsilon _{\mu \nu } F^{\mu \nu }  - A_0 J^0
.\label{vacu}
\end{equation} Since we are dealing with static fermions, we can
substitute $\partial^2$ by $-\partial^2_1$. In that case we write
\begin{equation}
{\cal L} =  - \frac{1}{4}F_{\mu \nu } \left( {1 - \frac{{e^2
}}{\pi }\frac{1}{{\partial _1^2  - 4\pi m\Sigma }}} \right)F^{\mu
\nu } - em\Sigma  \frac{1}{{\partial _1^2  - 4\pi m\Sigma }}\theta
- A_0 J^0 . \label{vacu2}
\end{equation}
But since the first term on the right-hand side (RHS) of
Eq. (\ref{vacu2}) gave as a result the potential given
in Eq. (\ref{lag}), we only need to consider the
second, $\theta$ dependent, term. Using (\ref{esc}), we may
write the expression for the scalar potential in the form
\begin{equation}
{\cal A}_0^\theta  (t,x) =   \int_0^1 {d\alpha x^1 } 2em\Sigma
\left( {\partial _1^2  - \lambda ^2 } \right)^{ - 1} \theta ,
\label{teta} \end{equation} where $\lambda ^2  \equiv \frac{{e^2
}}{\pi } + 4\pi m\Sigma$. If we now look at the limit of slowly
varying fields, we find that
\begin{equation}
{\cal A}_0^\theta  (t,x) = -  \int_0^1 {d\alpha \frac{{2em\Sigma
}}{{\lambda ^2 }}} x^1 \theta  =  - \frac{e}{{2\pi }}\left( {1 -
\frac{{\frac{{e^2 }}{\pi }}}{{\lambda ^2 }}} \right)x^1 \theta .
\label{teta2}
\end{equation}
Hence we see that the potential for two opposite charges ( for
$q=e$ ) at $y$ and $y^{\prime}$ is given by
\begin{equation}
V^\theta   =  - \frac{{e^2 }}{2}\left( {1 - \frac{{\frac{{e^2
}}{\pi }}}{{\lambda ^2 }}} \right)\frac{\theta }{\pi }|y -
y^\prime |.  \label{teta3}
\end{equation}
It is also, up to the $- \frac{\theta }{\pi }$ factor, just the
Coulomb potential. From this expression we see that, in the limit
of slowly varying fields, the inclusion of the $\theta$ term would
reduce the potential to
\begin{equation}
V = \frac{{e^2 }}{2}\left( {1 - \frac{{\frac{{e^2 }}{\pi
}}}{{\lambda ^2 }}} \right)\left( {1 - \frac{\theta }{\pi }}
\right)|y - y^\prime  | . \label{teta4}
\end{equation}
An immediate consequence of this is that for $\theta= \pi $ the
confinement term vanishes \cite{Elcio}.

\section{ACKNOWLEDGMENTS}

P.G. would like to thank J. Gamboa for the discussions.

\end{document}